\documentclass[useAMS,usenatbib]{mn2e}
\usepackage{mathptmx}
\usepackage{amssymb}

\usepackage{graphicx}
\usepackage[update,prepend]{epstopdf}
\newcommand{\hl}[1]{#1}
\newcommand{\hlrevtwo}[1]{#1}

\title{Is the Sulphur Anomaly in Planetary Nebulae Caused by the s-Process?}
\author[Luke J. Shingles and Amanda I. Karakas]{Luke J. Shingles\thanks{Email:
lukes@mso.anu.edu.au} and Amanda I. Karakas\\
Research School of Astronomy \& Astrophysics, Mount Stromlo Observatory, Weston Creek, ACT 2611, Australia}
\begin{document}

\date{Accepted 2013 February 27.  Received 2013 February 27; in original form 2012 December 27}

\pagerange{\pageref{firstpage}--\pageref{lastpage}} \pubyear{2013}

\maketitle

\label{firstpage}

\begin{abstract}
Motivated by unexplained observations of low sulphur abundances in planetary nebulae (PNe) and the PG1159 class of post asymptotic giant branch (AGB) stars, we investigate the possibility that sulphur may be destroyed by nucleosynthetic processes in low-to-intermediate mass stars during stellar evolution. We use a 3 M$_\odot$, $Z=0.01$ evolutionary sequence to examine the consequences of high and low reaction rate estimates of neutron captures onto sulphur and neighbouring elements. In addition we have also tested high and low rates for the neutron producing reactions $^{13}$C($\alpha$,n)$^{16}$O and $^{22}$Ne($\alpha$,n)$^{25}$Mg. We vary the mass width of a partially mixed zone (PMZ), which is responsible for the formation of a $^{13}$C pocket and is the site of the $^{13}$C($\alpha$,n)$^{16}$O neutron source. We test PMZ masses from zero up to an extreme upper limit of the entire He-intershell mass at $10^{-2}$ M$_\odot$. We find that the alternative reaction rates and variations to the partially mixed zone have almost no effect on surface sulphur abundances and do not reproduce the anomaly. \hl{To understand the effect of initial mass on our conclusions, 1.8 M$_\odot$ and 6 M$_\odot$ evolutionary sequences are also tested with similar results for sulphur abundances.} We are able to set a constraint on the size of the PMZ, as PMZ sizes that are greater than half of the He-intershell mass (in the 3 M$_\odot$ model) are excluded by comparison with neon abundances in planetary nebulae. We compare the 1.8 M$_\odot$ model's intershell abundances with observations of PG1159-035, whose surface abundances are thought to reflect the intershell composition of a progenitor AGB star. We find general agreement between the patterns of F, Ne, Si, P, and Fe abundances and a very large discrepancy for sulphur where our model predicts abundances that are 30-40 times higher than is observed in the star.
\end{abstract}

\section{Introduction}
After leaving the asymptotic giant branch (AGB), a post-AGB star may evolve to high temperatures ($>30,000$ K) on the timescale required to ionise the surrounding shell of ejected material and become visible as a planetary nebula (PN). As PNe are composed of envelope material from a progenitor AGB star, measurements of PN chemical abundances provide a way to test the predictions of AGB nucleosynthesis models \citep[e.g.,][]{Marigo:2003km,2009ApJ...690.1130K,2010A&A...517A..95P,2010PASA...27..227K}.

Many planetary nebulae (PNe) with approximately solar oxygen abundance ($\pm 0.4$ dex) have been found to have sulphur depletions of between 0.1 and 0.6 dex relative to the Sun \citep{Marigo:2003km}. A more detailed investigation with a larger sample of 85 PNe by \citet{Henry:2004ds} discovered that sulphur abundances in PNe are systematically lower than HII regions at the same metallicity, where metallicity in PNe is measured indirectly through the oxygen abundance. Specifically, \citet{Henry:2004ds} showed that the abundance trends between PNe and HII regions are co-linear in the Ne-O, Cl-O, and Ar-O planes, but are separated in the S-O plane, in which the trend-line of PNe is located below that of HII regions by 0.3 dex. The co-linear trends between Ne, Cl, Ar, and O, but not S single out sulphur as the anomalous element, and this has been labelled the `sulphur anomaly'. The sulphur anomaly has been independently confirmed by the observations of \citet{Milingo:2010er}.

\citet{Henry:2004ds} argued that because AGB models do not predict significant depletion of sulphur and sulphur does not readily condense into dust grains, the most likely cause of the sulphur anomaly is a failure to correctly account for sulphur in the highly ionised S$^{+3}$ state through the use of an Ionisation Correction Factor (ICF) and measurements of S$^{+1}$ and S$^{+2}$ abundances. However, infrared observations of PNe \citep[e.g.,][]{BernardSalas:2008kz} have directly measured S$^{+3}$ abundances using the [S IV] emission line at 10.5 $\mu$m. This was done without the need for an ICF and these observations show that the sulphur anomaly still exists and is in need of explanation.

If the observed low gas phase abundance of sulphur in PNe relative to the interstellar medium (as sampled by HII regions) reflects a decrease in elemental sulphur between the birth composition of a star and its surface layers at the final phases of stellar evolution, then an attractive solution would be to identify a nucleosynthetic process that is able to destroy sulphur during the intervening stages. The progenitors of planetary nebulae are typically low-mass stars \citep[e.g., 1.0 to 2.5 M$_\odot$;][]{2011A&A...531A..23P}, which evolve through the AGB phase and experience nucleosynthesis through H and He burning and the slow neutron capture process ($s$-process) \citep[e.g.,][]{1998ApJ...497..388G,1999ARA&A..37..239B}. During the AGB, nucleosynthesis products are periodically dredged up into the convective hydrogen-rich envelope as part of the thermal pulse cycle, so a depletion of sulphur in the He-intershell of an AGB star would result in a (smaller) depletion of sulphur at the stellar surface. A review of AGB evolution and modelling is given by \citet{2005ARA&A..43..435H}.

\citet{Henry:2012gd} provides an update on the status of the sulphur anomaly and discusses the still-viable explanations, including gas phase depletion due to dust or molecule formation, and the nuclear processing in AGB stars. Although it was argued that the sulphur anomaly is inconsistent with the predictions of existing nucleosynthesis models \citep[e.g.,][]{2010MNRAS.403.1413K}, there has not been an investigation into how modelling uncertainties such as nuclear reaction rates and the treatment of mixing affect predictions of surface sulphur abundances.

There is a separate physical site with unexplained sulphur depletion in stars of the type PG1159. PG1159 stars are extremely hot (75,000-200,000 K) post-AGB stars that are hydrogen deficient and helium rich, likely because of a late or very late helium shell flash that has consumed their remaining hydrogen envelope and exposed He-intershell material to the stellar surface \citep{2003ARA&A..41..391V,1991A&A...244..437W}. Although a very late thermal pulse and hydrogen ingestion episode may lead to some additional light element and $s$-process nucleosynthesis after the AGB \citep{2011ApJ...727...89H,2011ApJ...742..121S}, the resulting surface abundances are expected to largely reflect the intershell composition at the end of the AGB phase. With intershell matter at their surfaces, PG1159 stars provide a test of nucleosynthesis models that is relatively free of the uncertainties related to dredge-up efficiency that affect the surface abundances of AGB stars.

\citet{2006PASP..118..183W} report that PG1159 stars have highly scattered and generally low sulphur abundances ranging from 0.01-1 times solar, while the models of Herwig show He-intershell sulphur abundances at the end of the AGB that are 0.6-0.9 times solar. They suggest that a study is needed to understand how the uncertainties of neutron capture cross sections affect intershell abundances. \citet{2009Ap&SS.320..159W} interprets the discrepancy between low sulphur observations of PG1159 stars and sulphur preserving theoretical models as a failure of stellar modelling. If the current models' failure to reproduce the sulphur anomaly is an indication that our understanding of stellar nucleosynthesis is in need of refinement, then a solution to the sulphur problem may lead to a better understanding of other aspects of stellar nucleosynthesis, such as the mixing near convective boundaries and nuclear reaction rates.

\begin{figure}\label{fig:ncapturerates}
 \begin{center}\includegraphics[width=1.0\columnwidth]{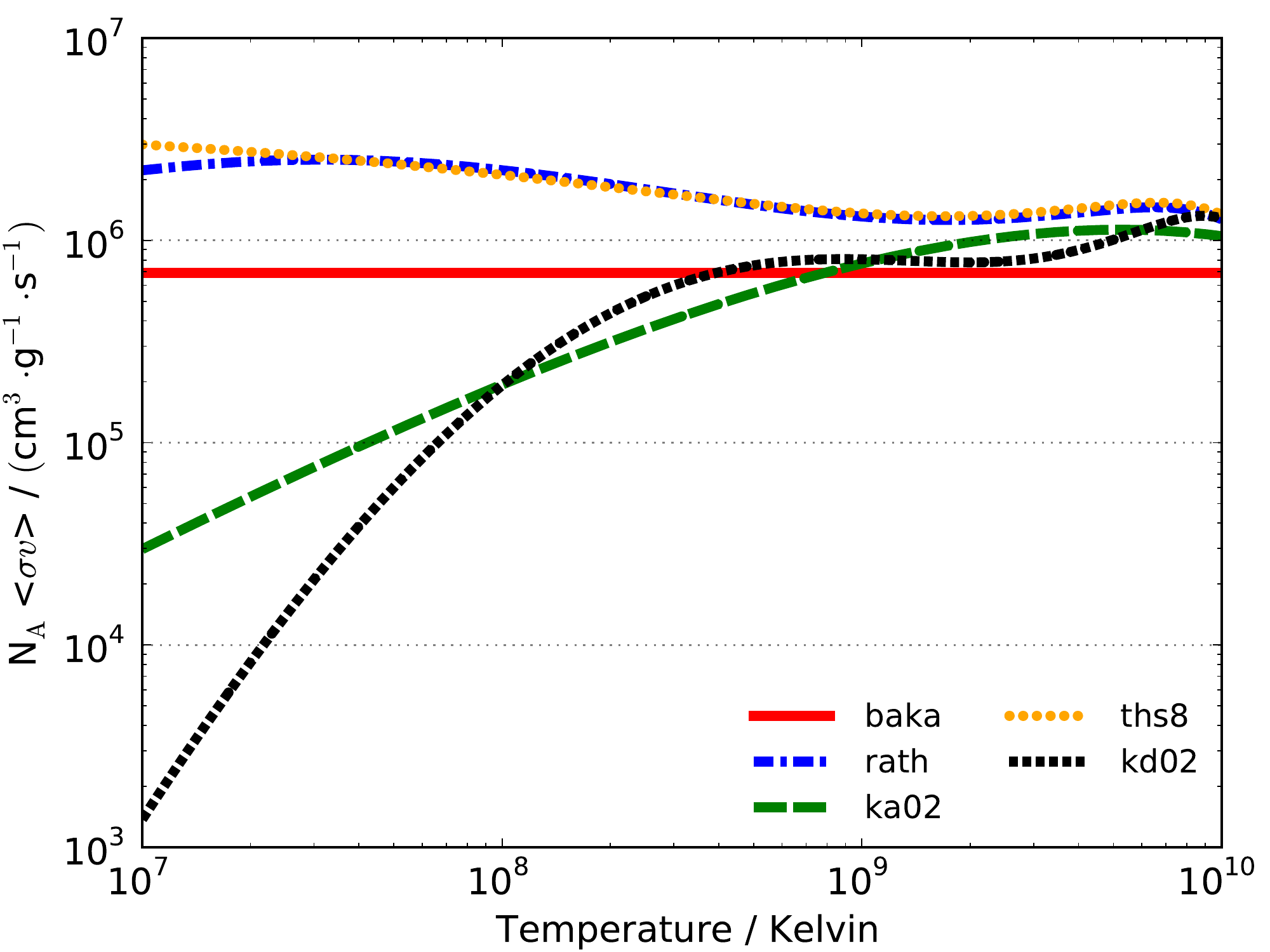}\end{center}
 \caption{Reduced reaction rates for $^{32}$S(n,$\gamma$)$^{33}$S as a function of temperature from several sources in the JINA Reaclib database \citep{Cyburt:2010ey}. Source labels are defined in Section \ref{sec:reactionratechanges}.}
\end{figure}

Sulphur has a significant nuclear charge ($Z=16$) that prevents it from strongly participating in charged-particle reactions (e.g., p- and $\alpha$-capture) at the temperatures and densities of AGB stellar interiors. For sulphur destruction, the much more likely pathway is neutron capture on S to produce the unstable isotopes $^{35}$S and $^{37}$S, which spontaneously decay via $\beta^-$ to $^{35}$Cl and $^{37}$Cl, respectively.  Although neutron-capture reactions with S are included in existing models, there is disagreement over the rates of these reactions that becomes particularly significant at temperatures below $10^9$ K. The disparity is evident in Figure \ref{fig:ncapturerates}, which shows the rate of the $^{32}$S(n,$\gamma$)$^{33}$S reaction as a function of temperature from several sources in the JINA Reaclib database \citep{Cyburt:2010ey}. Most neutron captures at the top of the He-intershell take place at a temperature of the order of $10^8$ K, and at this temperature the independently predicted rates disagree by up to factor of 10. At temperatures less than $10^7$ K, the neutron capture rates diverge rapidly and differ by over three orders of magnitude.

In this paper, we aim to determine whether the sulphur anomaly in PNe is the consequence of the nucleosynthetic processes in PN-progenitor AGB stars. To do this, we calculate models that span the range of relevant uncertainties in modelling low-mass stars -- neutron-capture reaction rates, neutron-producing reaction rates, and partial mixing zone profiles (which determine the size of the $^{13}$C-pockets) and compare their surface abundances to PNe observational data. We will examine the intershell abundances of our models in comparison with PG1159 observations and consider the significance that extra mixing (e.g., convective overshoot and rotation) could have reproducing the sulphur anomaly in PNe.

\section{Numerical Method \& Models}
We evolve our stellar evolutionary sequences from the zero-age main sequence to the tip of the AGB with the Mount Stromlo Stellar Structure Program, which has been updated to include C- and N-rich low temperature opacity tables from \citet{Lederer:2009da}, as described in \citet{2010ApJ...713..374K} and references therein. We use Reimers' formula \citep{Reimers:1975vw} with the parameter $\eta=0.4$ for mass loss during the first red giant branch and the \citet{1993ApJ...413..641V} prescription on the AGB.

The evolutionary code operates on a minimal set of nuclides that are involved in reactions that are highly exothermic (the pp-chains, CNO cycle, triple-$\alpha$, and $^{12}$C($\alpha$,$\gamma$)$^{16}$O reactions) and hence affect the stellar structure. The structure model generated by the evolutionary code is used as input to a post-process nucleosynthesis code. With time- and mass-dependent variables such as temperature, density, and the locations of convective boundaries defined in the structure model, the nucleosynthesis code recalculates abundances for a detailed network with time-dependent diffusive mixing for all convective zones \citep{Cannon:1993te}. 

\hl{As the energy generation in the He-flash convective zone is completely dominated by the triple-$\alpha$ reaction included in the structure model, we are able to modify the rates of weakly energetic or endothermic $\alpha$- and n-capture reactions and add a partially mixed zone (Section \ref{sec:pmz}) in the nucleosynthesis post-process without having to recalculate the stellar structure.} In this work, our detailed nuclear network consists of 125 species, which include many isotopes of P, S, and Cl to precisely account for the neutron-capture and $\beta$-decay reactions around sulphur.

Our chosen initial mass of 3.0 M$_\odot$ is near the upper end of PN progenitor masses. As shown in the results of \citet{2007PASA...24..103K}, 3.0 M$_\odot$ models experience a greater number of thermal pulses and third dredge-up than lower mass models, so this choice will exaggerate the effect of any possible sulphur depletion on surface abundances.

For our initial composition, we scale the solar abundances of \citet{2009ARA&A..47..481A} ($Z_\odot=0.0142$) such that the models' metallicity is $Z=0.704\,Z_\odot=0.01$. The models' final metallicity will be roughly a factor of two larger than the initial value (mostly due to the dredge-up of primary $^{12}$C during the thermally pulsing AGB phase), so the model will lie just above the centre of the PNe metallicity range of $0.3\,Z_\odot$ to $2\,Z_\odot$ \citep{2008ApJS..174..158S}. 

\subsection{Changes to Reaction Rates}\label{sec:reactionratechanges}
To explore the effect of rate uncertainties for neutron-capture reactions with sulphur and its nuclear neighbours, we select the two sources in the JINA Reaclib database that predict the highest and lowest rates for $^{32}$S(n,$\gamma$)$^{33}$S around the intershell temperature of $10^8$ K. Our standard rate case is the ReaclibV0.5 release by \cite{Cyburt:2010ey}, which, for this reaction includes experimental estimates from the KADoNiS database \citep{2006AIPC..819..123D} labelled `ka02'. The source that predicts the lowest rate, `kd02' is very similar to `ka02', except that fitting formulae have been adjusted to maintain accuracy at low temperatures. The highest rate source, `ths8' is comprised of theoretical estimates by Thomas Rauscher that were included as part of the REACLIB V1.0 release \citep{Cyburt:2010ey}. \hl{The current ReaclibV2.0 release has adopted `kd02' rates for the $^{32}$S(n,$\gamma$)$^{33}$S reaction.} Also included in Figure \ref{fig:ncapturerates} are the experimental rates of \citet{Bao:1987hy} (labelled `baka'), and the statistical model calculations by \citet{Rauch:2000gg} (`rath'), however, these rates have not been adopted in this study.

\subsection{Partial Mixing Zone}\label{sec:pmz}
The free neutrons available for the $s$-process in low-mass stars are primarily produced by $^{13}$C burning under radiative conditions via the $^{13}$C($\alpha$,n)$^{16}$O reaction \citep{1995ApJ...440L..85S,1998ApJ...497..388G}. Producing the seed $^{13}$C nuclei in stellar models requires the existence of a layer at the top of the $^{12}$C-rich intershell in which protons are `partially mixed' down from the envelope, thus enabling the CN cycle reaction $^{12}$C(p,$\gamma$)$^{14}$N($\beta^+$)$^{13}$C. The mixing process cannot be too efficient, or else the newly-created $^{13}$C nuclei will be destroyed by further proton capture to make $^{14}$N, which is a neutron poison, i.e., its large neutron-capture cross section significantly reduces the number of free neutrons available for the $s$-process. The physical mechanism behind the formation of a partially mixed zone (PMZ) is still a mystery, although some of the more likely possibilities include convective overshooting \citep{Herwig:2000ua,Cristallo:2004us}, rotational mixing \citep{2001MmSAI..72..277H}, or gravity-wave driven mixing \citep{Denissenkov:2003gx}.

\hl{Some models in the literature insert a $^{13}$C pocket directly at each thermal pulse, using a profile such as the \citet{1998ApJ...497..388G} standard (ST) case, which has a $^{13}$C pocket mass of $5 \times 10^{-4}$ M$_\odot$. \citet{2001MmSAI..72..277H} use a diffusive convective overshoot at the bottom of the envelope convection boundary with the parameter $f=0.016$ and find a $^{13}$C pocket width (where the $^{13}$C mass fraction is above $10^{-4}$) of about $2 \times 10^{-5}$ M$_\odot$ in a 3 M$_\odot$ model.

To experiment with different $^{13}$C pocket masses, the studies of \citet{Arlandini:1999eh} and \citet{Bisterzo:2010cd,MNR:MNR20670} insert a \citet{1998ApJ...497..388G} ST profile $^{13}$C (and $^{14}$N) pocket that has been scaled in $^{13}$C (and $^{14}$N) abundance. We instead scale the width (in mass coordinate) of an inserted proton profile, which not only controls the total mass of protons inserted (and the mass of the resulting $^{13}$C and $^{14}$N pockets) but also changes the radial position and extent over which the resulting neutron-captures take place. For a comparison involving both of these treatments of the $^{13}$C pocket, see the detailed discussion in \citet{2012ApJ...747....2L}.}

\begin{figure}
 \begin{center}\includegraphics[width=\columnwidth]{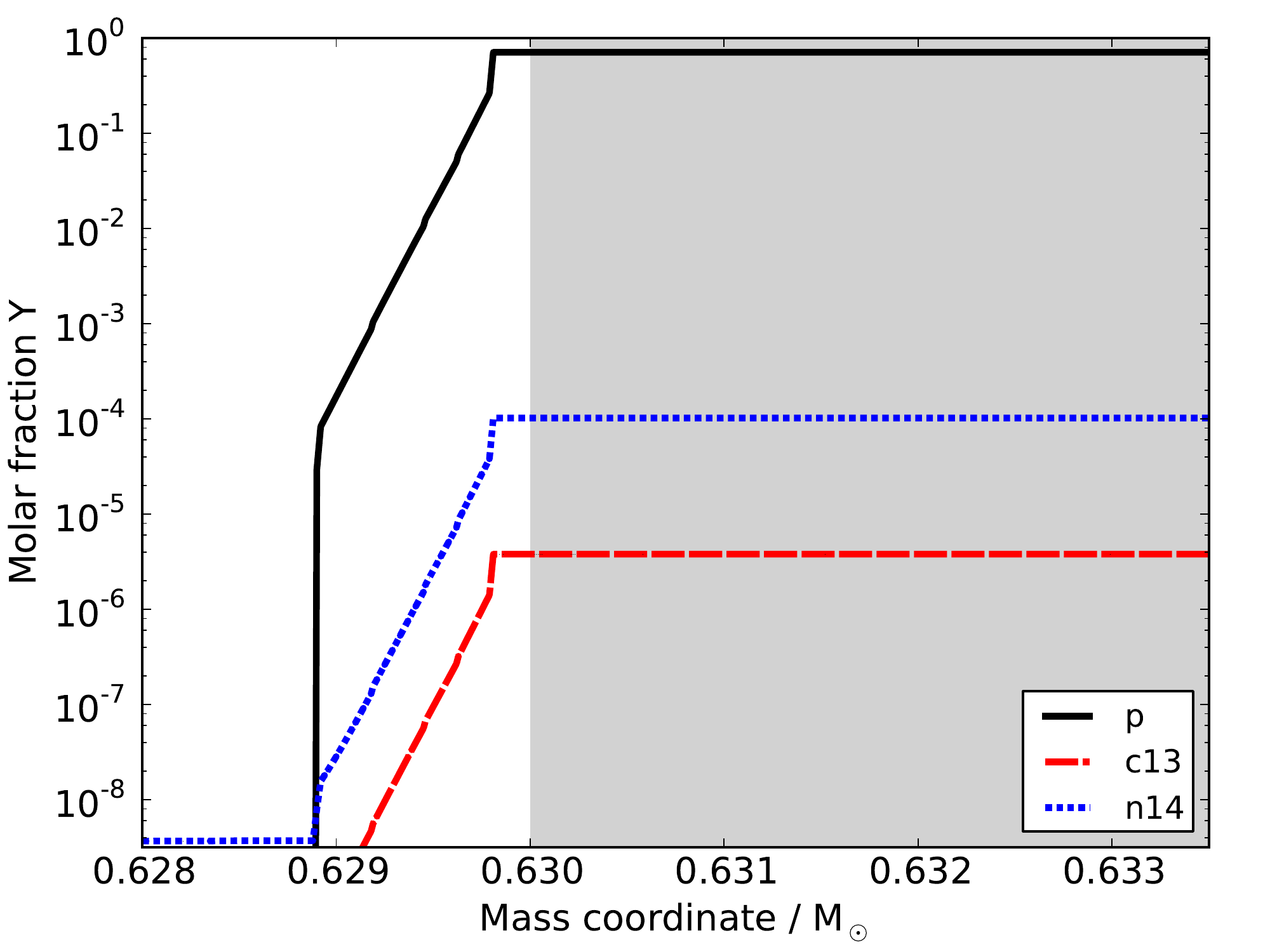}
 \includegraphics[width=\columnwidth]{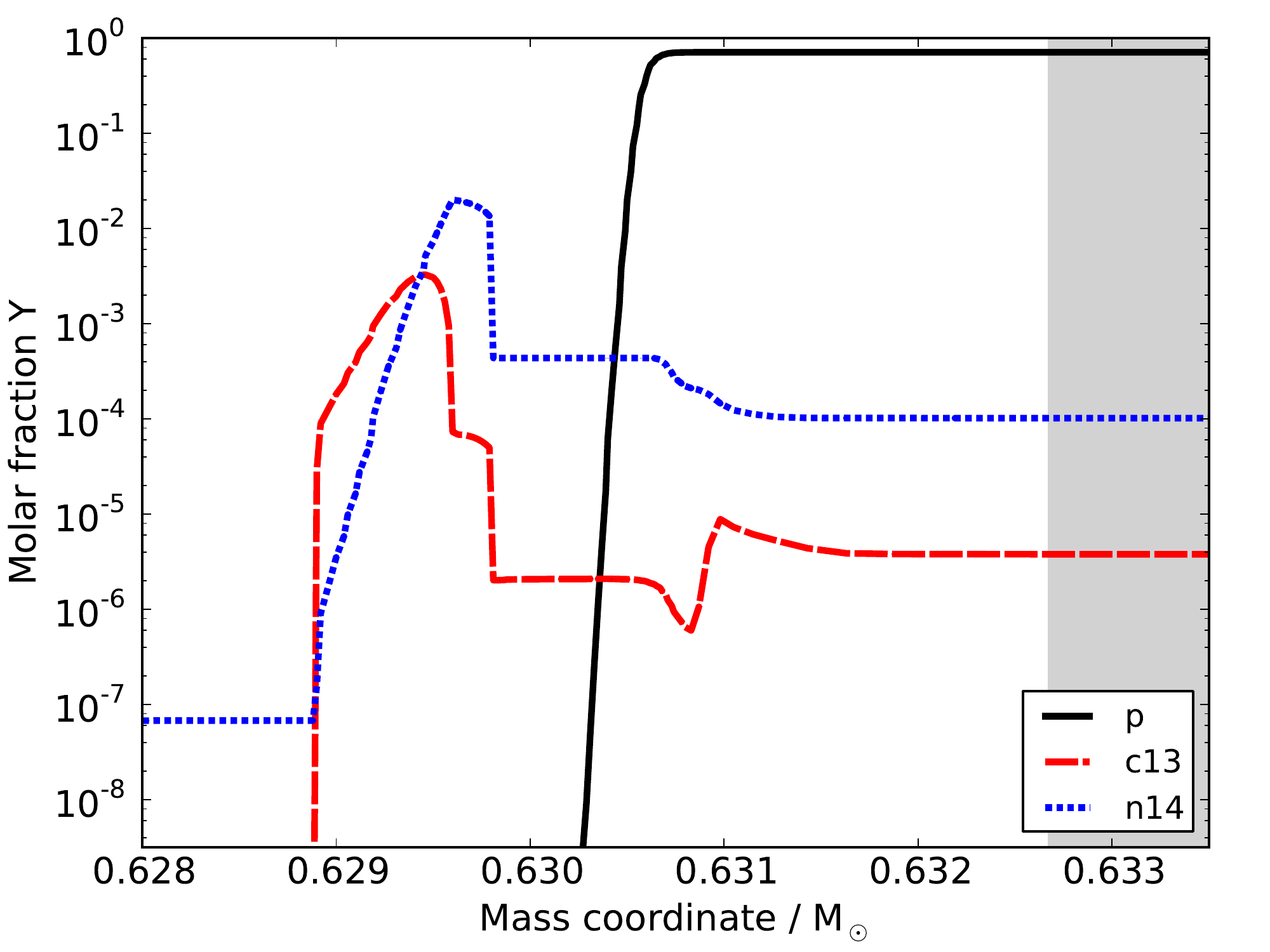}\end{center}
 \caption{Upper panel: The 3 M$_\odot$ model proton profile immediately after inserting a $10^{-3}$ M$_\odot$ partial mixing zone. Lower panel: As the envelope convection zone (shaded) retreats outwards in mass, proton capture reactions result in adjacent pockets of $^{13}$C and $^{14}$N forming near the top of the intershell. Y denotes the molar fraction, equal to (mass fraction) / (atomic mass).}\label{fig:c13n14pockets}
\end{figure}

\hl{Figure \ref{fig:c13n14pockets} shows the $^{13}$C and $^{14}$N pockets that form as a result of an exponential profile of protons inserted below the envelope convection zone. The protons are inserted identically after every thermal pulse with third dredge-up, at the time when the envelope convection zone reaches its deepest extent during a third dredge-up episode. The proton profile matches the envelope abundance at top of the PMZ and decreases exponentially to a mass fraction of $10^{-4}$ across a mass interval we refer to as the `PMZ mass'. This method is described in more detail in \citet{2004ApJ...615..934L} and is very similar to that used by \citet{2000A&A...362..599G}. 

To explore the results of exaggerated neutron capture nucleosynthesis, we test 3 M$_\odot$ models with PMZ masses of $(1,\,5,\mbox{ and }10) \times 10^{-3}$ M$_\odot$ in addition to a model with no partial mixing zone. Unless otherwise stated, we use our standard PMZ with a mass of $1\times 10^{-3}$ M$_\odot$. The PMZ mass of $10 \times 10^{-3}$ M$_\odot$ is included as an extreme upper limit for for the 3 M$_\odot$ model, as this profile spans the entire He-intershell and the partial mixing is not expected to penetrate into the degenerate C-O core.}

\section{Model Results}

\subsection{New Stellar Evolutionary Models}
The 1.8 M$_\odot$ model has previously been described in \citet{2010ApJ...713..374K}. We present new 3 M$_\odot$ and 6 M$_\odot$ sequences at $Z=0.01$ computed with Mount Stromlo evolutionary code.

\begin{table*}
  \centering
  \begin{minipage}{155mm}
  \caption{Structural and dredge-up parameters of the 3 M$_\odot$, $Z=0.01$ sequence for each thermal pulse cycle during the AGB. Columns are described in Section \ref{structure-labels}. \label{tbl:m3z01-structure}}
\begin{tabular}{l l l l l l l l l l l l l l}
	\hline
	TP	&	M$_{\mathrm{tot}}$	&	M$_{\mathrm{H}}$	&	$\Delta$M$_{\mathrm{DUP}}$	&	$\lambda$	&	T$_{\mathrm{He-shell}}$	&	T$_{\mathrm{H-shell}}$&	T$_{\mathrm{bce}}$		&	$\tau_{\mathrm{ip}}$	&	L$_{\mathrm{max}}$	&	R$_{\mathrm{max}}$	&	M$_{\mathrm{bol}}$	&	T$_\mathrm{eff}$	&	C/O\\
	\#	&	[M$_\odot$]	&	[M$_\odot$]	&	[$10^{-3}$ M$_\odot$]	&	\,	&	[$10^6$ K]	&	[$10^6$ K]	&	[$10^6$ K]	&	[$10^3$ yr]	&	[$10^3$ L$_\odot$]	&	[R$_\odot$]	&	[mag]	&	[K]	&	\,\\
\hline
1	&	2.99	&	0.607	&	0.00	&	0.00	&	181.6	&	53.0	&	2.3	&	0.00		&	3.5	&	146	&	-4.15	&	3797	&	0.36\\
2	&	2.99	&	0.610	&	0.00	&	0.00	&	214.1	&	57.3	&	2.7	&	58.27	&	5.0	&	188	&	-4.52	&	3775	&	0.36\\
3	&	2.99	&	0.614	&	0.00	&	0.00	&	226.3	&	58.9	&	2.8	&	77.52	&	5.6	&	204	&	-4.65	&	3749	&	0.37\\
4	&	2.99	&	0.619	&	0.00	&	0.00	&	238.2	&	60.7	&	3.0	&	84.12	&	6.3	&	223	&	-4.78	&	3715	&	0.37\\
5	&	2.99	&	0.625	&	0.00	&	0.00	&	245.6	&	62.0	&	3.2	&	85.53	&	6.9	&	238	&	-4.87	&	3682	&	0.37\\
6	&	2.99	&	0.631	&	0.90	&	0.14	&	253.0	&	63.1	&	3.3	&	83.52	&	7.4	&	251	&	-4.95	&	3580	&	0.37\\
7	&	2.99	&	0.637	&	2.42	&	0.34	&	261.7	&	64.3	&	3.5	&	82.65	&	7.9	&	265	&	-5.03	&	3520	&	0.39\\
8	&	2.99	&	0.642	&	3.71	&	0.47	&	269.7	&	65.4	&	3.8	&	81.83	&	8.5	&	278	&	-5.10	&	3474	&	0.48\\
9	&	2.99	&	0.647	&	5.24	&	0.61	&	273.0	&	66.2	&	4.0	&	81.51	&	9.0	&	290	&	-5.16	&	3437	&	0.63\\
10	&	2.99	&	0.651	&	6.57	&	0.70	&	278.9	&	66.8	&	4.3	&	82.36	&	9.5	&	303	&	-5.22	&	3410	&	0.84\\
11	&	2.99	&	0.655	&	7.32	&	0.72	&	283.1	&	67.3	&	4.6	&	82.72	&	9.9	&	313	&	-5.27	&	3387	&	1.07\\
12	&	2.99	&	0.658	&	8.01	&	0.76	&	287.2	&	67.6	&	5.0	&	81.37	&	10.3	&	318	&	-5.31	&	3377	&	1.30\\
13	&	2.98	&	0.661	&	8.49	&	0.77	&	290.8	&	67.8	&	4.9	&	80.67	&	10.7	&	352	&	-5.35	&	3258	&	1.53\\
14	&	2.98	&	0.664	&	8.77	&	0.78	&	294.0	&	68.0	&	5.1	&	78.92	&	11.0	&	371	&	-5.38	&	3198	&	1.78\\
15	&	2.98	&	0.666	&	8.81	&	0.78	&	296.5	&	68.2	&	5.2	&	76.91	&	11.3	&	391	&	-5.41	&	3145	&	2.03\\
16	&	2.98	&	0.669	&	9.00	&	0.80	&	298.6	&	68.3	&	5.3	&	74.18	&	11.6	&	409	&	-5.44	&	3097	&	2.27\\
17	&	2.97	&	0.671	&	8.89	&	0.78	&	300.8	&	68.5	&	0.0	&	72.56	&	11.8	&	426	&	-5.46	&	3055	&	2.50\\
18	&	2.96	&	0.673	&	8.83	&	0.80	&	301.7	&	68.6	&	5.5	&	69.27	&	12.1	&	444	&	-5.48	&	3019	&	2.74\\
19	&	2.92	&	0.676	&	8.99	&	0.81	&	302.8	&	68.7	&	5.6	&	67.58	&	12.3	&	460	&	-5.50	&	2983	&	2.98\\
20	&	2.83	&	0.678	&	8.98	&	0.81	&	304.9	&	68.8	&	5.6	&	65.74	&	12.5	&	482	&	-5.52	&	2948	&	3.23\\
21	&	2.17	&	0.680	&	7.85	&	0.71	&	305.8	&	68.8	&	5.4	&	63.95	&	12.6	&	581	&	-5.53	&	2901	&	3.48\\
22	&	1.23	&	0.682	&	7.09	&	0.72	&	303.6	&	68.4	&	3.4	&	58.91	&	12.5	&	843	&	-5.52	&	2633	&	3.79\\
\hline
\end{tabular}
\end{minipage}
\end{table*}

\begin{table*}
  \centering
  \begin{minipage}{155mm}
  \caption{Structural and dredge-up parameters of the 6 M$_\odot$, $Z=0.01$ sequence for each thermal pulse cycle during the AGB. Columns are described in Section \ref{structure-labels}. \label{tbl:m6z01-structure}}
\begin{tabular}{l l l l l l l l l l l l l l}
	\hline
	TP	&	M$_{\mathrm{tot}}$	&	M$_{\mathrm{H}}$	&	$\Delta$M$_{\mathrm{DUP}}$	&	$\lambda$	&	T$_{\mathrm{He-shell}}$	&	T$_{\mathrm{H-shell}}$&	T$_{\mathrm{bce}}$		&	$\tau_{\mathrm{ip}}$	&	L$_{\mathrm{max}}$	&	R$_{\mathrm{max}}$	&	M$_{\mathrm{bol}}$	&	T$_\mathrm{eff}$	&	C/O\\
	\#	&	[M$_\odot$]	&	[M$_\odot$]	&	[$10^{-3}$ M$_\odot$]	&	\,	&	[$10^6$ K]	&	[$10^6$ K]	&	[$10^6$ K]	&	[$10^3$ yr]	&	[$10^3$ L$_\odot$]	&	[R$_\odot$]	&	[mag]	&	[K]	&	\,\\
\hline
1	&	5.95	&	0.911	&	0.00	&	0.00	&	236.7	&	76.9	&	31.4	&	0.00	&	23.1	&	436	&	-6.19	&	3518	&	0.37\\
2	&	5.95	&	0.912	&	0.07	&	0.09	&	247.3	&	78.1	&	38.6	&	3.23	&	24.0	&	449	&	-6.23	&	3494	&	0.37\\
3	&	5.95	&	0.913	&	0.28	&	0.30	&	256.5	&	79.3	&	46.9	&	3.36	&	25.0	&	462	&	-6.27	&	3486	&	0.37\\
4	&	5.95	&	0.914	&	0.54	&	0.51	&	264.9	&	80.4	&	56.0	&	3.44	&	26.2	&	478	&	-6.32	&	3475	&	0.37\\
5	&	5.95	&	0.914	&	0.71	&	0.60	&	272.7	&	81.4	&	62.7	&	3.53	&	27.6	&	498	&	-6.38	&	3465	&	0.38\\
6	&	5.95	&	0.915	&	0.93	&	0.73	&	279.6	&	82.2	&	66.6	&	3.65	&	29.0	&	519	&	-6.44	&	3457	&	0.38\\
7	&	5.95	&	0.915	&	1.10	&	0.79	&	286.7	&	83.0	&	69.4	&	3.77	&	30.4	&	537	&	-6.49	&	3449	&	0.38\\
8	&	5.95	&	0.916	&	1.26	&	0.84	&	293.0	&	83.7	&	71.6	&	3.92	&	31.7	&	554	&	-6.53	&	3446	&	0.35\\
9	&	5.95	&	0.916	&	1.41	&	0.87	&	299.2	&	84.4	&	73.5	&	4.10	&	32.7	&	569	&	-6.57	&	3426	&	0.31\\
10	&	5.95	&	0.917	&	1.56	&	0.90	&	304.8	&	85.0	&	75.3	&	4.28	&	33.6	&	580	&	-6.60	&	3398	&	0.25\\
11	&	5.95	&	0.917	&	1.56	&	0.84	&	310.3	&	85.7	&	77.0	&	4.49	&	34.3	&	590	&	-6.62	&	3376	&	0.18\\
12	&	5.95	&	0.917	&	1.65	&	0.88	&	313.8	&	86.2	&	78.5	&	4.55	&	34.7	&	596	&	-6.63	&	3347	&	0.13\\
13	&	5.94	&	0.917	&	1.83	&	0.90	&	319.6	&	86.9	&	80.0	&	4.85	&	35.4	&	603	&	-6.65	&	3344	&	0.09\\
14	&	5.94	&	0.918	&	1.95	&	0.93	&	323.6	&	87.5	&	81.2	&	4.99	&	35.7	&	609	&	-6.66	&	3326	&	0.07\\
15	&	5.94	&	0.918	&	2.06	&	0.93	&	328.0	&	88.0	&	82.1	&	5.19	&	36.3	&	616	&	-6.68	&	3319	&	0.06\\
16	&	5.93	&	0.918	&	2.11	&	0.92	&	330.6	&	88.3	&	82.8	&	5.30	&	36.9	&	625	&	-6.70	&	3308	&	0.06\\
17	&	5.93	&	0.918	&	2.19	&	0.93	&	334.7	&	88.6	&	83.2	&	5.42	&	37.4	&	631	&	-6.71	&	3294	&	0.06\\
18	&	5.92	&	0.919	&	2.25	&	0.93	&	338.0	&	88.9	&	83.6	&	5.52	&	37.9	&	638	&	-6.73	&	3288	&	0.06\\
19	&	5.91	&	0.919	&	2.30	&	0.93	&	340.4	&	89.1	&	83.9	&	5.62	&	38.4	&	644	&	-6.74	&	3278	&	0.06\\
20	&	5.90	&	0.919	&	2.36	&	0.94	&	342.8	&	89.3	&	84.1	&	5.69	&	38.8	&	651	&	-6.75	&	3269	&	0.06\\
21	&	5.88	&	0.919	&	2.41	&	0.94	&	344.7	&	89.4	&	84.4	&	5.79	&	39.2	&	657	&	-6.76	&	3265	&	0.06\\
22	&	5.86	&	0.920	&	2.21	&	0.84	&	346.1	&	89.5	&	84.5	&	5.86	&	39.6	&	663	&	-6.77	&	3257	&	0.06\\
23	&	5.84	&	0.920	&	2.22	&	0.84	&	347.4	&	89.6	&	84.6	&	5.92	&	39.9	&	668	&	-6.78	&	3251	&	0.06\\
24	&	5.80	&	0.920	&	2.54	&	0.95	&	347.5	&	89.7	&	84.7	&	5.94	&	40.2	&	673	&	-6.79	&	3244	&	0.07\\
25	&	5.75	&	0.920	&	2.56	&	0.94	&	348.5	&	89.8	&	84.8	&	6.06	&	40.4	&	678	&	-6.80	&	3237	&	0.07\\
26	&	5.69	&	0.920	&	2.57	&	0.94	&	350.8	&	89.8	&	84.8	&	6.06	&	40.5	&	682	&	-6.80	&	3228	&	0.07\\
27	&	5.60	&	0.921	&	2.35	&	0.86	&	354.2	&	89.8	&	84.7	&	6.05	&	40.5	&	686	&	-6.80	&	3216	&	0.07\\
28	&	5.46	&	0.921	&	2.45	&	0.89	&	350.2	&	89.7	&	84.5	&	6.04	&	40.4	&	690	&	-6.80	&	3205	&	0.08\\
29	&	5.17	&	0.921	&	2.63	&	0.95	&	352.0	&	89.6	&	84.1	&	6.10	&	40.1	&	695	&	-6.79	&	3193	&	0.08\\
30	&	4.85	&	0.921	&	2.67	&	0.94	&	352.3	&	89.4	&	83.3	&	6.25	&	39.3	&	700	&	-6.77	&	3169	&	0.08\\
31	&	4.54	&	0.921	&	2.69	&	0.94	&	348.0	&	89.0	&	82.4	&	6.27	&	38.0	&	704	&	-6.73	&	3137	&	0.09\\
32	&	4.23	&	0.921	&	2.72	&	0.94	&	351.0	&	88.7	&	81.2	&	6.32	&	36.7	&	709	&	-6.69	&	3104	&	0.11\\
33	&	3.94	&	0.922	&	2.75	&	0.95	&	347.9	&	88.2	&	79.5	&	6.38	&	35.4	&	712	&	-6.65	&	3072	&	0.13\\
34	&	3.64	&	0.922	&	2.80	&	0.95	&	354.1	&	87.8	&	77.4	&	6.51	&	34.0	&	715	&	-6.61	&	3042	&	0.15\\
35	&	3.36	&	0.922	&	2.86	&	0.95	&	351.3	&	87.3	&	74.3	&	6.65	&	32.6	&	717	&	-6.56	&	3011	&	0.20\\
36	&	3.08	&	0.922	&	2.90	&	0.94	&	352.5	&	86.7	&	69.1	&	6.91	&	31.1	&	720	&	-6.51	&	2981	&	0.29\\
37	&	2.79	&	0.923	&	2.97	&	0.95	&	354.6	&	86.0	&	57.5	&	7.24	&	29.6	&	730	&	-6.46	&	2944	&	0.41\\
38	&	2.50	&	0.923	&	3.03	&	0.95	&	352.9	&	85.3	&	39.2	&	7.67	&	28.5	&	754	&	-6.42	&	2901	&	0.56\\
39	&	2.21	&	0.923	&	3.05	&	0.93	&	357.7	&	84.7	&	22.3	&	7.99	&	27.7	&	783	&	-6.39	&	2856	&	0.74\\
40	&	1.92	&	0.923	&	3.11	&	0.97	&	359.2	&	84.1	&	6.7	&	7.92	&	27.2	&	809	&	-6.37	&	2817	&	0.95\\
\hline
\end{tabular}
\end{minipage}
\end{table*}

\label{structure-labels}Table \ref{tbl:m3z01-structure} and Table \ref{tbl:m6z01-structure} list the structural parameters of each thermal pulse (TP) in the 3 M$_\odot$ and 6 M$_\odot$ models, respectively. The parameters include the total mass $($M$_{\mathrm{tot}})$ and the H-exhausted core mass $($M$_{\mathrm{H}})$ measured at the beginning of the TP, the mass dredged-up into the envelope $(\Delta$M$_{\mathrm{DUP}})$ and the dredge-up efficiency parameter $(\lambda=\Delta\mathrm{M}_{\mathrm{DUP}}/\Delta \mathrm{M}_{\mathrm{H}}$, where $\Delta$ denotes the change between the previous and current TP$)$ measured after the TP. Also included are the maximum temperatures of the He- and H-burning shells $($T$_{\mathrm{He-shell}}$, T$_{\mathrm{H-shell}})$ during the TP. \hl{In the interpulse period $(\tau_{\mathrm{ip}})$, we sample the maximum temperature at the base of the convective envelope $($T$_{\mathrm{bce}})$, the minimum luminosity $($L$_{\mathrm{max}})$, the maximum stellar radius (R$_{\mathrm{max}}$), the minimum bolometric magnitude $($M$_{\mathrm{bol}})$, the minimum effective temperature $($T$_{\mathrm{eff}})$, and the carbon-to-oxygen number ratio (C/O).}

In our 3 M$_\odot$ model, the temperature at the base of the convective envelope never exceeds $5.6 \times 10^6$ K, so there is no proton capture nucleosynthesis in the envelope (hot-bottom burning), which requires temperatures above about $(40\mbox{--}50) \times 10^6$ K. Without active CN-cycling in the envelope, third dredge-up raises the surface C/O ratio, eventually causing the model to become carbon-rich (C/O $>$1) after the 11th thermal pulse.

Envelope temperatures in the 6 M$_\odot$ model are in the range $(31\mbox{--}85) \times 10^6$ K, high enough that hot-bottom burning is active and prevents the surface from becoming carbon-rich up to the last (40th) thermal pulse, at which time the C/O ratio is 0.95. At the end of the AGB, the C/O ratio has increased to 1.16.

The maximum temperature in the helium burning shell generally increases with thermal pulse number in the 3 M$_\odot$ model, reaching a temperature of about 300 $\times 10^6$ K after the 15th thermal pulse. Above about $300\times 10^6$ K, the $^{22}$Ne($\alpha$,n)$^{25}$Mg reaction starts to become active, producing a burst of neutrons at the base of the intershell during convective thermal pulses. The $^{22}$Ne neutron source operates in addition to the $^{13}$C source, so this indicates that the intershell matter will be subject to additional neutron captures over the last few thermal pulses.

\begin{figure}
 \begin{center}\includegraphics[width=\columnwidth]{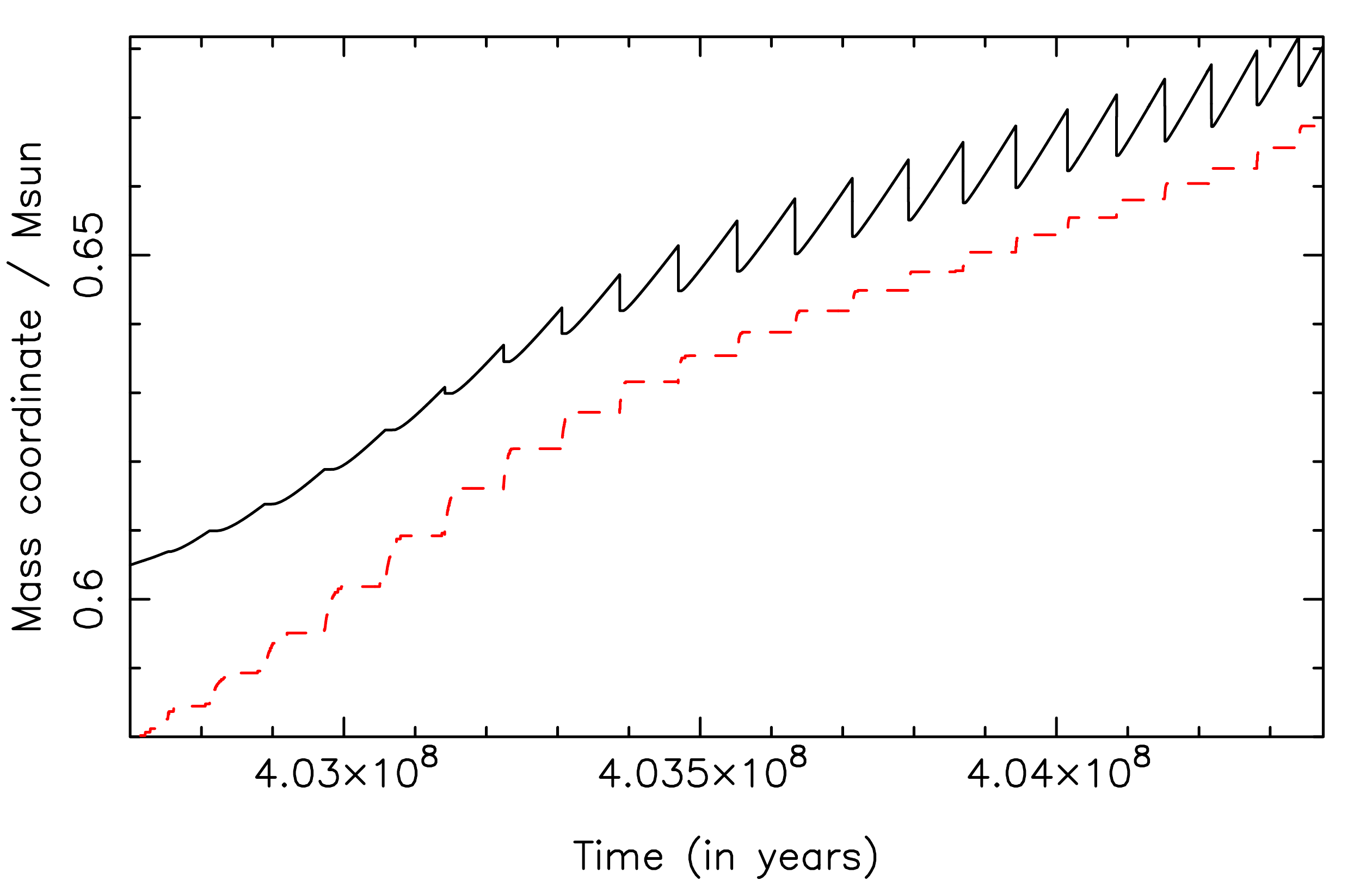}\end{center}
 \caption{Mass of the He-exhausted (dashed) and H-exhausted (solid) cores during the thermally pulsing AGB phase as a function of time in the 3 M$_\odot$ model.}\label{fig:m3z01-core-masses}
\end{figure}

The 3 M$_\odot$ model experiences a total of 22 thermal pulses before the AGB phase is terminated by mass loss at a final core mass of 0.68 M$_\odot$ (Figure \ref{fig:m3z01-core-masses}). At this time, the total amount of intershell matter that has been dredged up into the envelope is 0.120 M$_\odot$. With an envelope mass of about 2 M$_\odot$ for most of the AGB, there is enough material dredged up that the results of neutron capture nucleosynthesis in the He-intershell will be evident in the surface abundances of the model. This amount of dredge-up is a factor of a few compared to our evolutionary models with initial masses of 1.8 M$_\odot$ and 6 M$_\odot$, which dredge up a total of 0.041 M$_\odot$ and 0.082 M$_\odot$ of intershell matter, respectively. 

The cumulative dredge-up quantity of 0.12 M$_\odot$ in our 3 M$_\odot$, $Z=0.01$ model is comparable to other models of the same mass and similar metallicity. \citet{Lugaro:2003ew} compare 3 M$_\odot$, $Z=0.02$ models (the same mass but higher metallicity than our model) that have been computed independently with the Mount Stromlo stellar structure program, FRANEC \citep{1998ApJ...497..388G}, and EVOL \citep[as used by][with hydrodynamic overshoot included at all convective boundaries]{Herwig:2000ua}, finding cumulative dredge-up quantities of 0.08 M$_\odot$, 0.044 M$_\odot$, and 0.10 M$_\odot$, respectively. A 3 M$_\odot$, $Z=0.02$ model calculated with the Cambridge STARS code, which computes mixing and burning as a single step, dredges up a total of 0.13 M$_\odot$ after 20 thermal pulses in \citet{2004MNRAS.352..984S}. 

\subsection{Nucleosynthesis Model Results}\label{sec:nucmodelresults}

\begin{table*}
  \centering
  \begin{minipage}{125mm}
  \caption{Log $\epsilon^*$ surface elemental abundances, carbon-to-oxygen ratio (by number) and $^{12}$C/$^{13}$C isotopic ratios at the end of the AGB in our 3 M$_\odot$, $Z=0.01$ model.}\label{tbl:rates-surface-results}
\begin{tabular}{l c c c c c c c c r}
		          	 	& Ne   		& Mg   		& Si			&	S    		& P    		& Cl   		& Ar 		&	C/O	&	$^{12}$C/$^{13}$C  \\
\hline
scaled solar initial ($Z=0.01$)		&	7.802	&	7.472	&	7.382	&	6.992	&	5.242	&	5.062	&6.272	&	0.550	&	89.4391\\
\hline
m3z01-standard-pmz1	&	8.266	&	7.546	&	7.416	&	7.024	&	5.326	&	5.133	&6.302	&	4.279	&	305.146\\
m3z01-kd02-pmz1		&	8.267	&	7.547	&	7.416	&	7.023	&	5.318	&	5.134	&6.302	&	4.279	&	305.329\\
m3z01-ths8-pmz1		&	8.265	&	7.545	&	7.417	&	7.022	&	5.313	&	5.152	&6.302	&	4.280	&	305.278\\
\hline
m3z01-ths8-pmz0		&	8.167	&	7.531	&	7.417	&	7.023	&	5.280	&	5.136	&6.302	&	4.549	&	318.118\\
m3z01-ths8-pmz5		&	8.502	&	7.602	&	7.418	&	7.020	&	5.391	&	5.179	&6.301	&	3.467	&	259.238\\
m3z01-ths8-pmz10		&	8.627	&	7.640	&	7.419	&	7.021	&	5.407	&	5.186	&6.300	&	2.874	&	219.838\\
\hline
\end{tabular}
$^* \log\epsilon(X) = \log\left(N_X/N_H\right) + 12$
\end{minipage}
\end{table*}

\begin{figure}
 \begin{center}\includegraphics[width=\columnwidth]{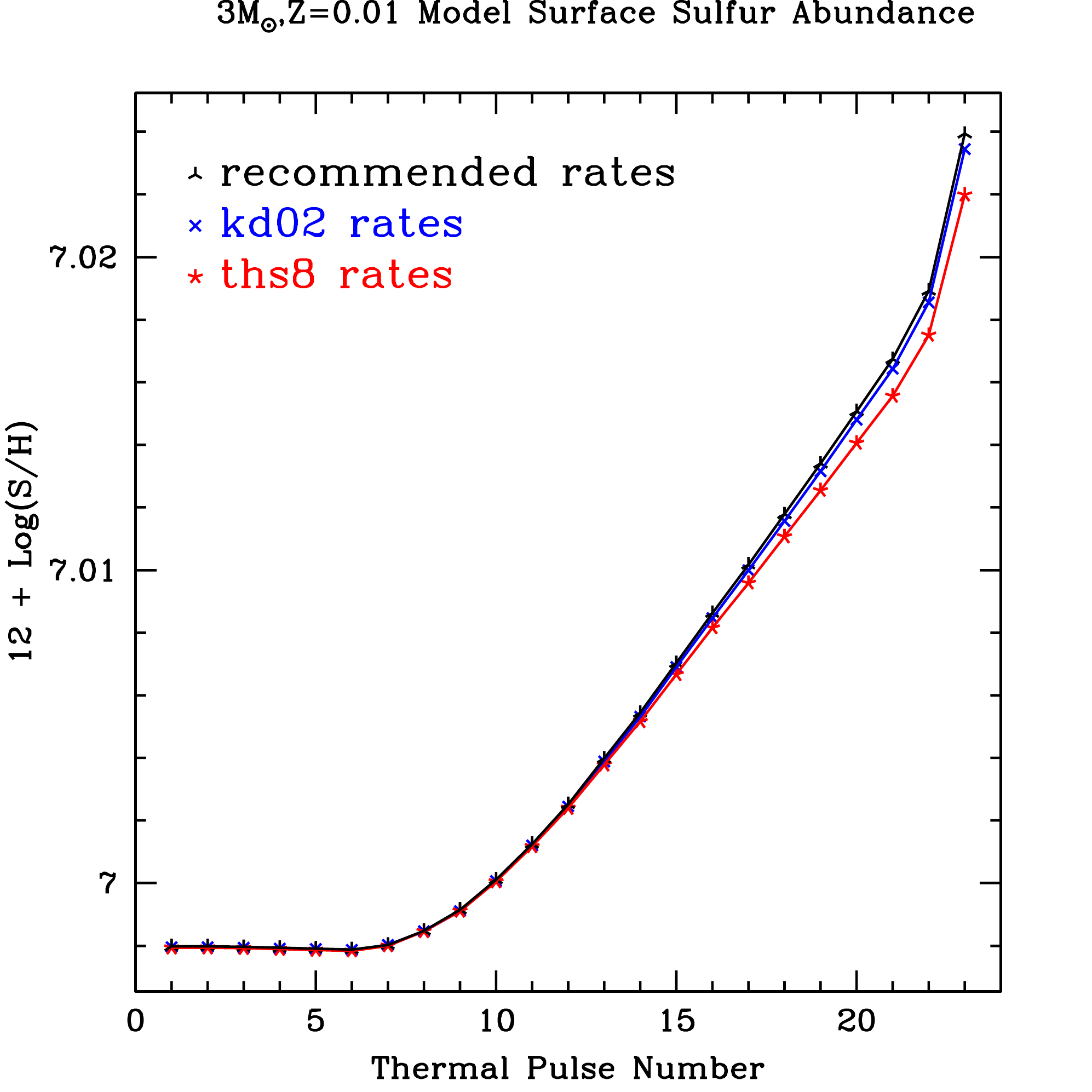}\end{center}
 \caption{Surface sulphur abundance as a function of thermal pulse number during the AGB phase in the 3 M$_\odot$, $Z=0.01$ model with a PMZ size of $10^{-3}$ M$_\odot$ and three neutron-capture rate sources.}\label{fig:rate-results-s}
\end{figure}

Table \ref{tbl:rates-surface-results} presents the final surface abundance results of the 3 M$_\odot$ model with standard (`ka02') and alternative neutron capture rates (`kd02' and `ths8') and several partial mixing zone sizes (0, 1, 5, and 10 $\times 10^{-3}$~M$_\odot$), as illustrated in Figure \ref{fig:rate-results-s}. The alternative neutron capture rates are seen to have little effect ($<0.001$ dex) on the surface abundance of sulphur at the end of the AGB phase. The variations to the partial mixing zone size also leave sulphur abundances virtually unchanged. In each of our cases, the surface abundance of sulphur is not depleted but instead shows a very slight increase ($<0.03$ dex) from pre-main sequence to the end of the AGB (Figure \ref{fig:rate-results-s}). 

\begin{figure}
 \begin{center}\includegraphics[width=\columnwidth,height=0.75\columnwidth]{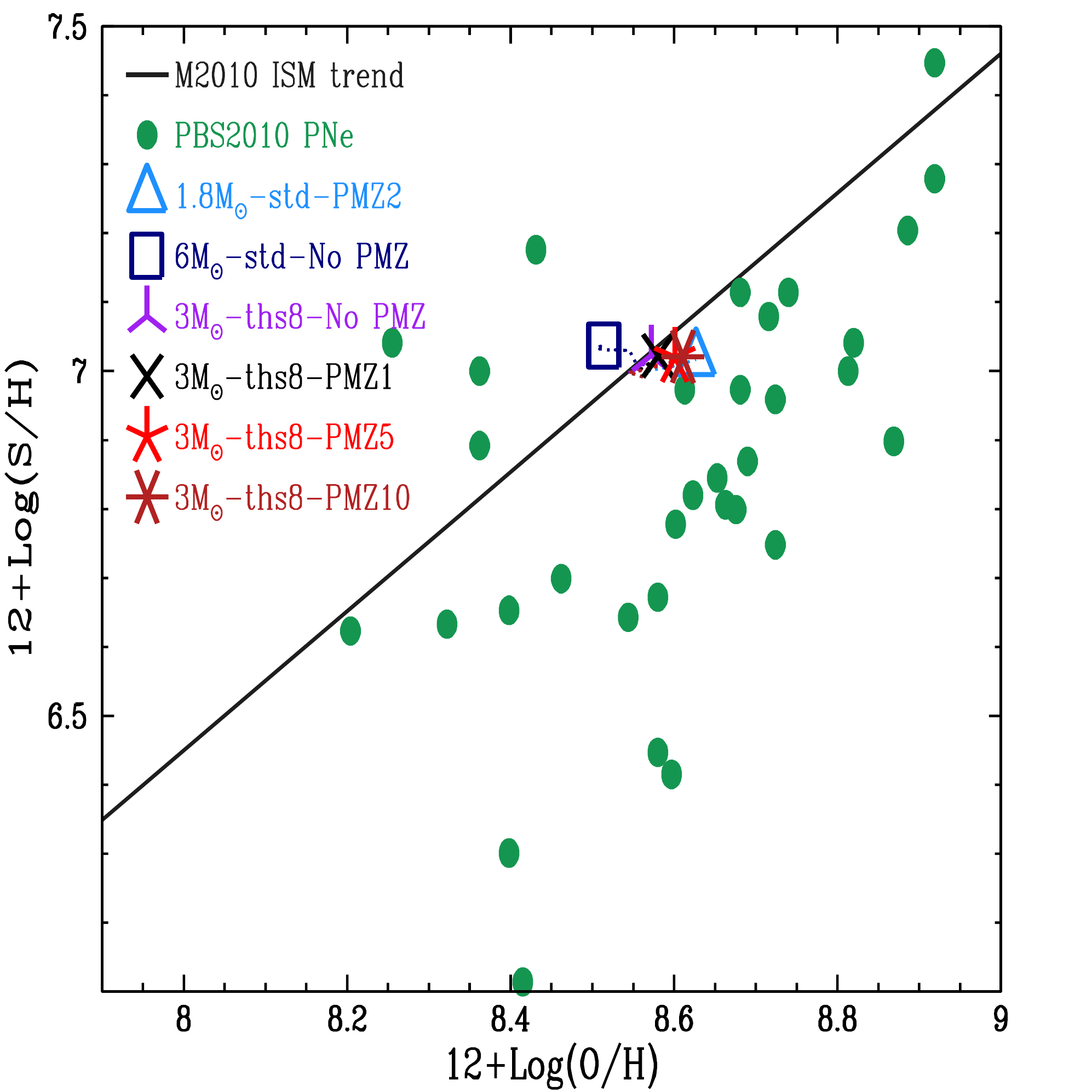}\end{center}
 \caption{Surface abundance results in the S versus O plane for 3 M$_\odot$ models with PMZ sizes of (1, 5, and 10)$\times 10^{-3}$ M$_\odot$, the 1.8M$_\odot$ model with a $2\times10^{-3}$ PMZ, and the 6 M$_\odot$ with no PMZ. Included for comparison are the PNe observational data of \citet{2010A&A...517A..95P} and the interstellar medium (ISM) trend of \citet{Milingo:2010er} from observations of HII regions and blue compact galaxies.}\label{fig:surfresults-svso}
\end{figure}

\begin{figure}
 \begin{center}\includegraphics[width=\columnwidth,height=0.75\columnwidth]{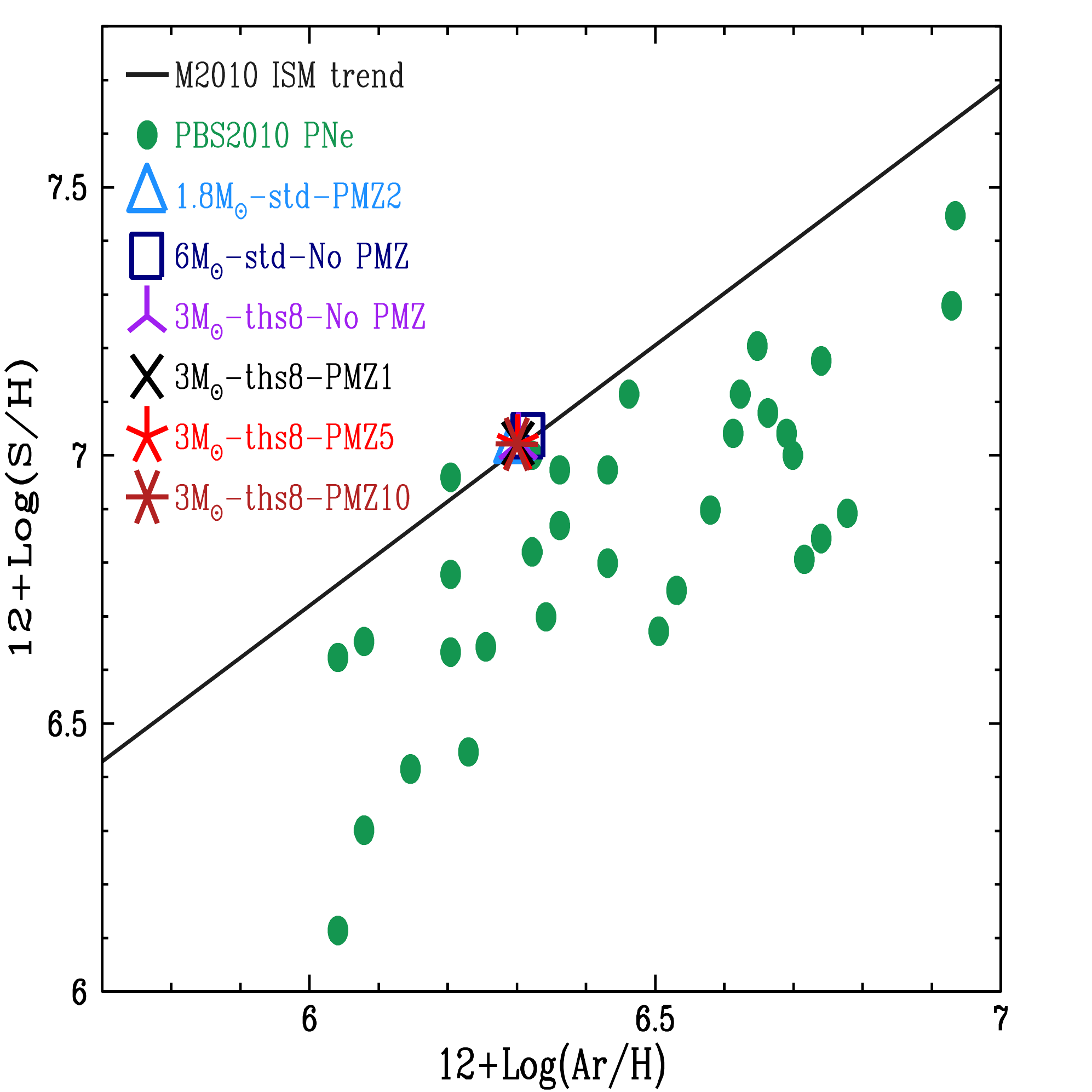}\end{center}
 \caption{Same as Figure \ref{fig:surfresults-svso} but for S versus Ar.}\label{fig:surfresults-svsar}
\end{figure}

In Figure \ref{fig:surfresults-svso}, we show surface abundance evolution in the S vs. O plane for 3 M$_\odot$ models with a range of PMZ sizes, as well as the 1.8 and 6 M$_\odot$ models for comparison with observational PNe abundances from \citet{2010A&A...517A..95P}. \hl{Although it is unlikely for a 6 M$_\odot$ star to produce a detectable planetary nebula, the 6 M$_\odot$ abundances are included to show the extent to which our results are dependent on the stellar initial mass.} An enhancement in O by up to 0.2 dex is seen in the 1.8 and 3 M$_\odot$ models, but this is not enough to explain the observational trend of PNe. Figure \ref{fig:surfresults-svsar} showing the results in the S versus Ar plane confirms that neither S nor Ar are significantly processed during low-mass evolution, and that none of the models can account for the trend of low S abundances in PNe.

\begin{figure}
 \begin{center}\includegraphics[width=\columnwidth]{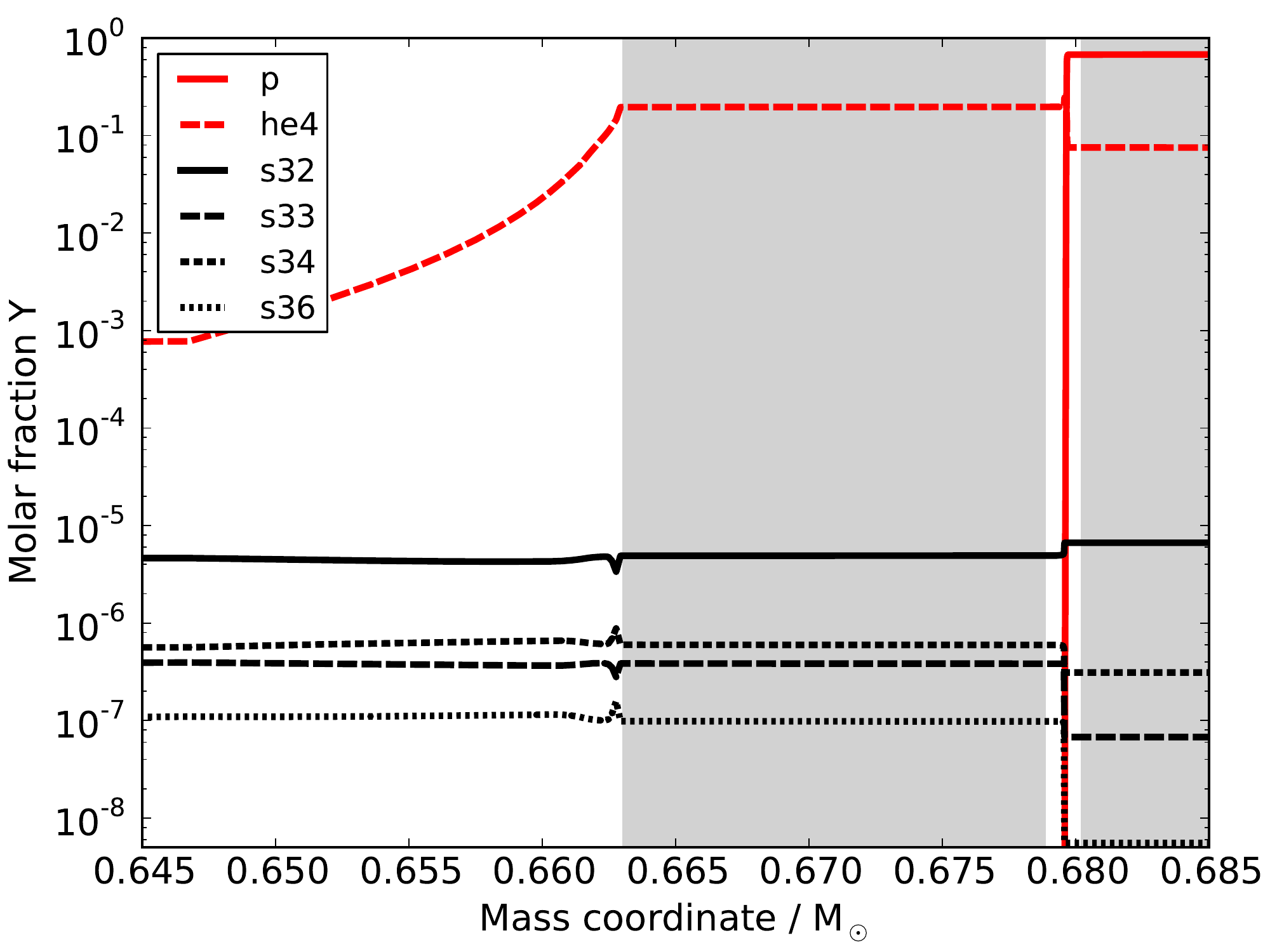}\end{center}
 \caption{Abundances in the He-intershell after the second last thermal pulse of the 3 M$_\odot$, $Z=0.01$ model with `ths8' rates and a PMZ mass of $1\times10^{-3}$ M$_\odot$. The shaded regions indicate convective zones.}\label{fig:composition}
\end{figure}

The sulphur abundance not only changes very little at the surface, but remains relatively constant throughout the entire stellar interior at the end of the AGB. The abundance discontinuity in (Figure \ref{fig:composition}) at the intershell-envelope boundary shows the results of a small $^{32}$S depletion in the intershell by conversion into more neutron-rich isotopes, indicated by increases in $^{33}$S, $^{34}$S, $^{35}$S, and $^{36}$S abundances. The increase in $^{36}$S abundance indicates that neutron captures onto unstable $^{35}$S nuclei occur on a timescale comparable to its $\beta^+$-decay mean lifetime of 126.3 days \citep{Audi:2003id}. \hl{The sulphur depletion (by 22\% or 0.1 dex) in the intershell is too small to resolve the sulphur anomaly in PNe, which requires sulphur depletions of typically 0.3 dex (and up to 0.6 dex) in the hydrogen-rich envelope.}

\begin{figure}
 \begin{center}\includegraphics[width=\columnwidth,height=0.75\columnwidth]{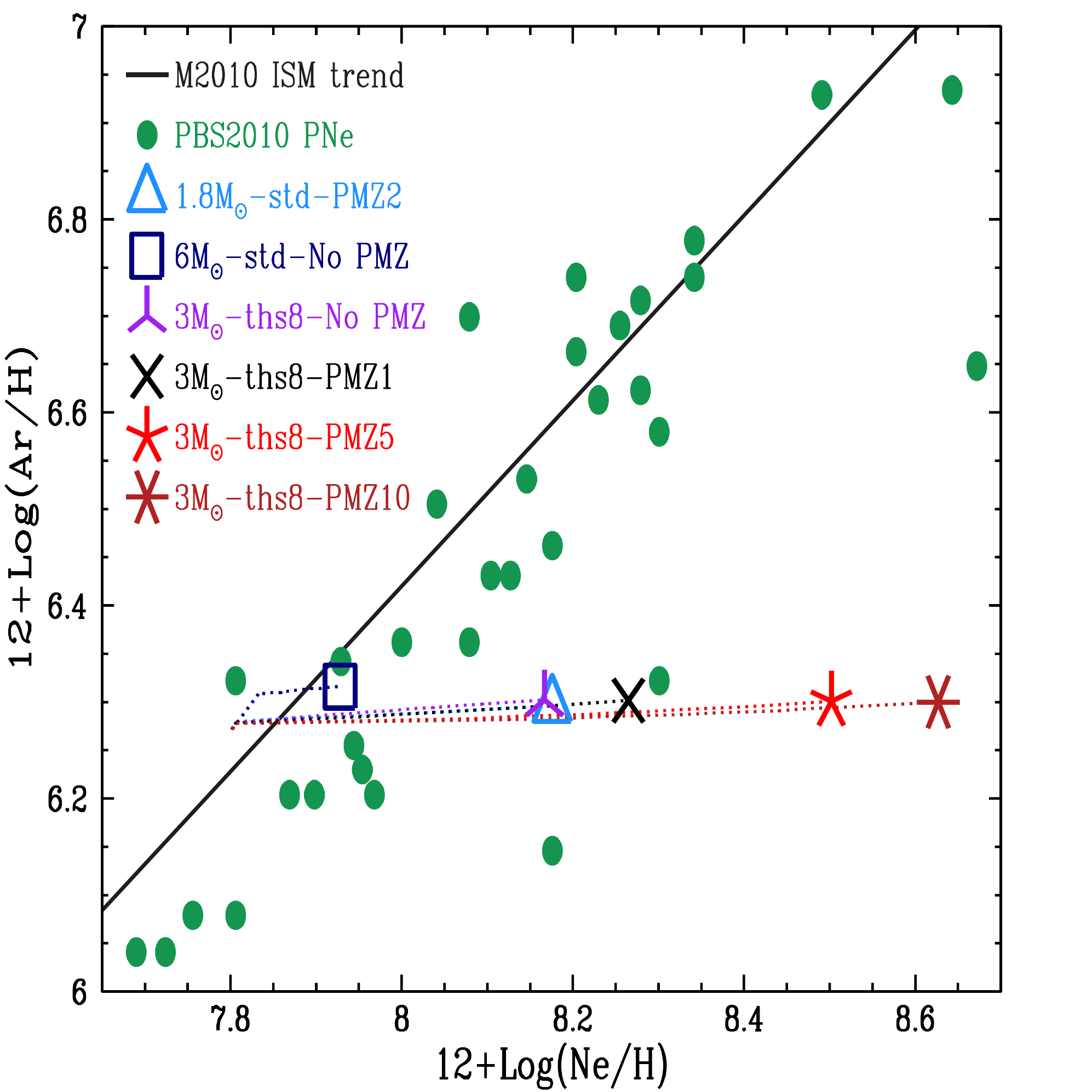}\end{center}
 \caption{Same as Figure \ref{fig:surfresults-svso} but for Ar versus Ne.}\label{fig:surfresults-arvsne}
\end{figure}
From the results in Table \ref{tbl:rates-surface-results}, we see that the neon surface abundances increase significantly with increases to the mass of the partial mixing zone. The elemental increase in neon is due to $^{22}$Ne production from primary $^{14}$N in convective pulses via the reaction chain $^{14}$N($\alpha$,$\gamma$)$^{18}$F($\beta^+$)$^{18}$O($\alpha$,$\gamma$)$^{22}$Ne. The size of the PMZ correlates with the size of the resulting $^{14}$N pocket, which adds to the quantity of $^{14}$N available for the production of $^{22}$Ne, as shown in Figure \ref{fig:surfresults-arvsne}.

The position of our models with large PMZ sizes in the Ar versus Ne plane is far from the observational PNe data (Figure \ref{fig:surfresults-arvsne}) and from this we conclude that the partial mixing zone widths larger than $5 \times 10^{-3}$ M$_\odot$ (50\% of the He-intershell mass in the 3 M$_\odot$ model) are excluded by the observations.

\begin{table*}
  \centering
  \begin{minipage}{110mm}
  \caption{Final log $\epsilon^*$ surface abundances for 3 M$_\odot$, $Z=0.01$ model with alternative rate sources: $^{13}$C($\alpha$,n)$^{16}$O low and high estimates from \citet[][NACRE collaboration]{1999NuPhA.656....3A} and $^{22}$Ne($\alpha$,n)$^{25}$Mg low, median, high, and high$\times600$ rate tables from \citet{2010NuPhA.841...31I}. Our standard case for comparison uses the ReaclibV0.5 release, which includes fits to the NACRE adopted rates for both reactions.}\label{tbl:neutron-production-surface-results}
\begin{tabular}{l c c c c c r}
		         	 						& Ne   	& Mg   	& S    	& P    	& Cl   	& Ar\\
\hline
Initial			 						& 7.802	& 7.472	& 6.992	& 5.242	& 5.062	& 6.272\\
Standard	(JINA ReaclibV0.5)			  		& 8.266	& 7.546	& 7.024	& 5.326	& 5.133	& 6.302\\
$^{13}$C($\alpha$,n)$^{16}$O NACRE-low 		& 8.200	& 7.537	& 7.024	& 5.325	& 5.135	& 6.302\\
$^{13}$C($\alpha$,n)$^{16}$O NACRE-high		& 8.200	& 7.537	& 7.024	& 5.323	& 5.134	& 6.303\\
$^{22}$Ne($\alpha$,n)$^{25}$Mg Iliadis2010-low 	& 8.207	& 7.516	& 7.024	& 5.319	& 5.128	& 6.303\\
$^{22}$Ne($\alpha$,n)$^{25}$Mg Iliadis2010-med 	& 8.206	& 7.517	& 7.024	& 5.320	& 5.130	& 6.303\\
$^{22}$Ne($\alpha$,n)$^{25}$Mg Iliadis2010-high	& 8.205	& 7.517	& 7.024	& 5.320	& 5.130	& 6.303\\
$^{22}$Ne($\alpha$,n)$^{25}$Mg Iliadis2010-high-x600	& 7.868	&	8.068	&	7.017	&	5.686	&	5.146	&	6.295\\
\hline
\end{tabular}
* $\log\epsilon(X) = \log\left(N_X/N_H\right) + 12$
\end{minipage}
\end{table*}

We have measured the effect of rate uncertainties on the two most important neutron-producing reactions by independently testing the high and low rates for $^{13}$C($\alpha$,n)$^{16}$O from \citet[][NACRE collaboration]{1999NuPhA.656....3A} and high, low, and adopted rates for $^{22}$Ne($\alpha$,n)$^{25}$Mg from \citet{2010NuPhA.841...31I}. Table \ref{tbl:neutron-production-surface-results} shows that these alternative rates have little effect on sulphur.

As with most charged particle reactions, the rate of the $^{22}$Ne($\alpha$,n)$^{25}$Mg reaction is very highly temperature-sensitive. As an extreme example, if the He-burning shell temperature was significantly higher by 33\% (0.4~GK\footnote{where 1~GK  $=1\times 10^{9}$ K.} instead of 0.3~GK), then the \citet{2010NuPhA.841...31I} high rate predicts a reaction rate increase from $3.01 \times 10^{-11}$ to $1.80 \times 10^{-8}$ cm$^3\cdot$g$^{-1}\cdot$s$^{-1}$, i.e., a factor of 631. To simulate the effect of a highly increased He-shell temperature on neutron production, we test a model with $^{22}$Ne($\alpha$,n)$^{25}$Mg rates boosted by a factor of 600. The surface abundance results of the boosted reaction rates are listed in Table \ref{tbl:neutron-production-surface-results} and include: a reduction in Ne (0.3 dex), increase in Mg (0.5 dex), an increase in P (0.4 dex) and a small decrease in S (0.01 dex). The implications of these results discussed in Section \ref{sec:discussion}.

\subsection{Comparison to PG1159-035}

\begin{table*}
  \centering
  \begin{minipage}{165mm}
  \caption{Top: Observed surface abundances of PG1159 stars PG1159-035 and PG1144+005 with the Sun for comparison.
\newline Bottom: Intershell abundance results of the models, measured during intershell convection at the last or second last thermal pulse. Rate definitions are given in the text.
\newline References: (1) \citet{2009ARA&A..47..481A}, (2) \citet{2007A&A...462..281J}, (3) \citet{2011A&A...531A.146W}, (4) \citet{2006PASP..118..183W}}\label{tbl:intershell-observations}
\begin{tabular}{l c c c c c c c c c c c}
\hline
\multicolumn{3}{l}{{\bf Stars}}					& 				& \multicolumn{8}{l}{{\bf Mass Fractions}}\\
 & & 	Metallicity Z									& Mass					&	C	& O		& F		& Ne	& Si		& P		& S		& Fe		\\
 & & 												& [M$_\odot$]				&		&		&[$10^{-7}$]&[$10^{-2}$]&[$10^{-4}$]&[$10^{-5}$]&[$10^{-4}$]&[$10^{-4}$]\\
\hline
\multicolumn{2}{l}{Solar$^{(1)}$}	& 0.014				& 1.00					& 0.003	& 0.006	& 3.66	& 0.14	& 7.3	& 0.58	& 3.4	& 14\\
\multicolumn{2}{l}{Scaled Solar}	& 0.010				& -						& 0.002	& 0.004	& 2.44	& 0.09	& 4.9	& 0.39	& 2.3	& 9.5\\ 
\multicolumn{2}{l}{PG1144+005$^{(4)}$}& -				& 0.60					& 0.570	& 0.016	& 100	& 2.00	& -		& -		& -		&  -	\\
\multicolumn{2}{l}{PG1159-035$^{(2,3)}$}& $\simeq Z_\odot$&0.536$_{-0.010}^{+0.068}$	& 0.480	& 0.170	& 32.0	& 2.00	& 3.6	& 0.64	& 0.05	& 13	\\
\hline
\multicolumn{3}{l}{{\bf Models $\mathbf{(Z=0.01)}$}}			&	& \multicolumn{8}{l}{{\bf He-Intershell Mass Fractions}}\\
M$_{\mathrm{initial}}$&Rates	& M$_{\mathrm{PMZ}}$	&M$_{\mathrm{f}}$				& C		& O		& F		& Ne	& Si		& P		& S		& Fe	\\
[0ex][M$_\odot$]		&			& [M$_\odot$]			&[M$_\odot$]				&		&		&[$10^{-7}$]&[$10^{-2}$]&[$10^{-4}$]&[$10^{-5}$]&[$10^{-4}$]&[$10^{-4}$]\\
\hline
1.8			& ka02		& $2\times10^{-3}$			& 0.59					& 0.167	& 0.012	& 202	& 2.19	& 5.0	& 2.2	& 2.2	& 8.3\\
3.0			& kd02		& $1\times10^{-3}$			& 0.68					& 0.176	& 0.004	& 667	& 3.39	& 5.2	& 1.1	& 2.1	& 8.8\\
3.0			& ka02		& $1\times10^{-3}$			& 0.68					& 0.175	& 0.004	& 657	& 3.39	& 5.1	& 1.2	& 2.2	& 8.9\\
3.0			& ths8		& No PMZ				& 0.68					& 0.183	& 0.002	& 547	& 2.56	& 5.3	& 5.2	& 2.0	& 9.4\\
3.0			& ths8		& $1\times10^{-3}$			& 0.68					& 0.175	& 0.004	& 657	& 3.39	& 5.4	& 1.0	& 1.9	& 9.0\\
3.0			& ths8		& $5\times10^{-3}$ 			& 0.68					& 0.146	& 0.006	& 809	& 6.26	& 5.6	& 2.4	& 1.8	& 7.8\\
3.0			& ths8		& $10\times10^{-3}$		& 0.68					& 0.123	& 0.008	& 847	& 8.32	& 5.9	& 2.3	& 1.9	& 8.0\\
3.0			& $^{22}$Ne-Il10-high-x600	& $1\times10^{-3}$	& 0.68				& 0.179	& 0.004	& 2700	& 0.18	& 12		& 13		& 1.6	& 1.6\\
6.0			& ka02		& No PMZ				& 0.98					& 0.201	& 0.004	& 65.1	& 1.22	& 9.8	& 3.5	& 1.7	& 4.4\\
\hline
\end{tabular}
\end{minipage}
\end{table*}

\begin{figure}
 \begin{center}\includegraphics[width=\columnwidth]{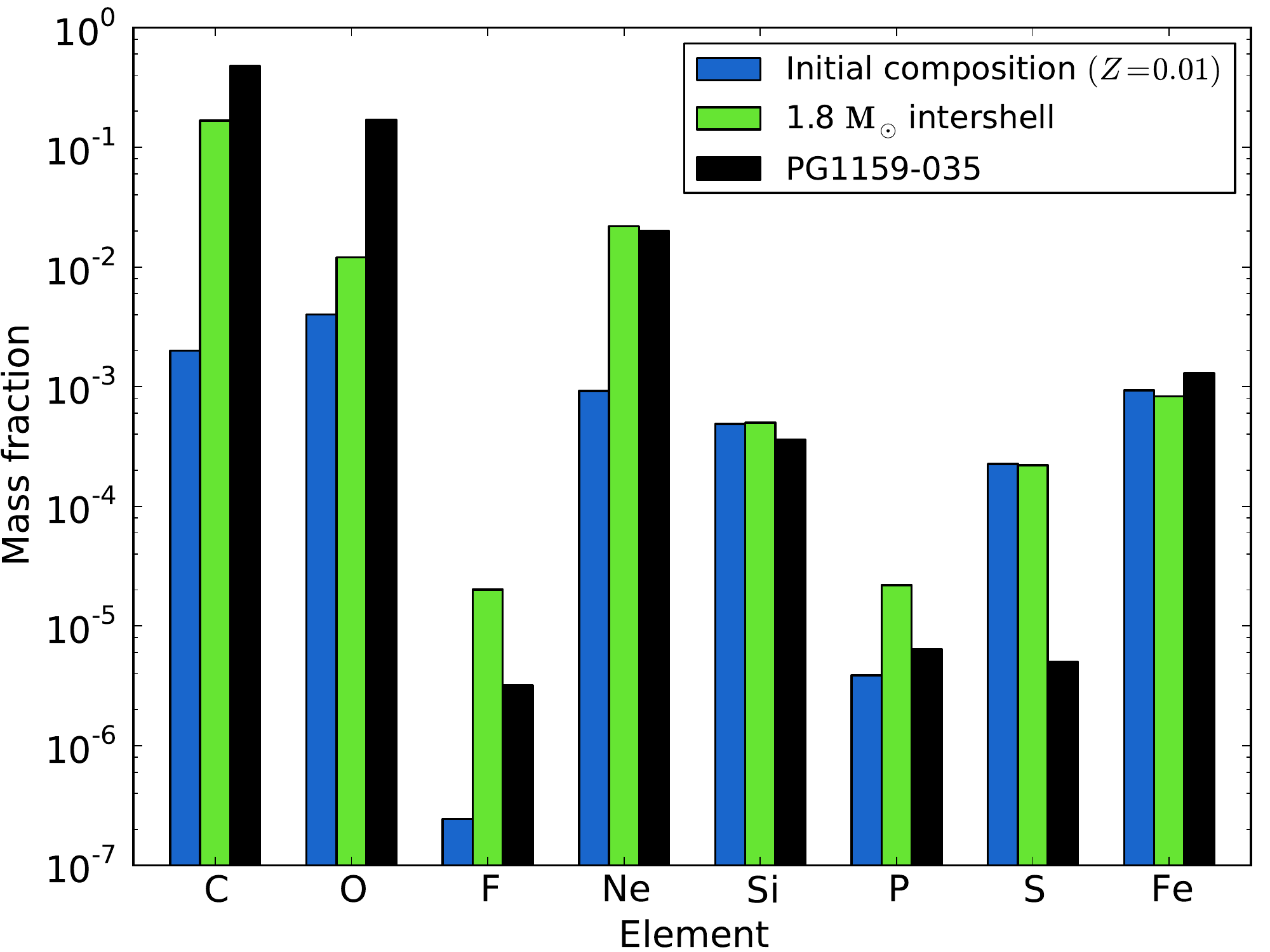}\end{center}
 \caption{Comparison of our 1.8 M$_\odot$ intershell abundances with surface abundances of PG1159-035 from \citet{2007A&A...462..281J} and \citet{2011A&A...531A.146W}.}\label{fig:pg1159035}
\end{figure}

\hlrevtwo{The hydrogen-deficient and helium-rich surface chemistry of PG1159 stars is likely caused by a late helium shell flash and consequent convection zone that extends into the hydrogen-rich surface layers \citep{Schoenberner:1979vv,1983ApJ...264..605I,1999A&A...349L...5H}. \hl{The nucleosynthesis during post-AGB evolution is expected to be a relatively small addition to the nucleosynthesis during the preceding $\sim 20$ thermal pulses \citep{1999A&A...349L...5H}, with the exception of a very late helium shell flash and proton ingestion episode, which may significantly affect the surface composition of light elements (up to oxygen) and heavy neutron-capture elements \citep{2011ApJ...727...89H,2011ApJ...742..121S}. Hence, although our stellar models terminate at the tip of the AGB, we expect that our intershell abundances of elements heavier than oxygen and much lighter than iron will approximately match the surface abundances of PG1159 stars.}} Abundance measurements for elements heavier than oxygen in PG1159 stars are rare in the literature. However, abundances in the prototype PG1159-035 have been measured for many elements including Si, P, S, and Fe \citep{2007A&A...462..281J,2011A&A...531A.146W} and we include them in Table \ref{tbl:intershell-observations}. Also shown in Table \ref{tbl:intershell-observations} are the intershell abundances of our 1.8 M$_\odot$, 3.0 M$_\odot$, and 6.0 M$_\odot$ models for comparison.

\citet{1991A&A...244..437W} estimated the mass of PG1159-035 at $0.605^{+.13}_{-.04}$ M$_\odot$ by comparison with the evolutionary tracks of \citet{Schoenberner:1979vv}. The mass measured by \citet{1991A&A...244..437W} is a close match to our 1.8 M$_\odot$ model, which has a mass at the end of the AGB of 0.59 M$_\odot$. However, based on more recent evolutionary calculations by \citet{MillerBertolami:2006ko}, \citet{2011A&A...531A.146W} recently revised the mass of PG1159-059 downwards to $0.536^{+.068}_{-.010}$ M$_\odot$. The predicted final mass of our 1.8 M$_\odot$ model is still contained within the uncertainties of the revised measurement, although a better fit may be achieved by a model with a lower initial mass.

In Figure \ref{fig:pg1159035}, we plot the abundances of PG1159-035 together with the intershell abundances of our 1.8 M$_\odot$ model. The closely matching Fe abundances indicate that our model has a similar initial metallicity to PG1159-035, as low mass stellar evolution leaves Fe abundances almost unchanged. We find that our model overproduces fluorine, with a resulting intershell mass fraction that is 2.0 times that of PG1159-035. This may be caused by our model having a higher initial mass than PG1159-035.

\hlrevtwo{Our models do not include convective overshoot, and we therefore find carbon and oxygen abundances that are too low to match the surface of PG1159-035 \citep{1991A&A...244..437W}. The high carbon and oxygen abundances of the PG1159 stars have been shown to be consistent with overshoot into the CO core \citep{1999A&A...349L...5H}, whereas our abundance results match other standard models of AGB stars without convective overshoot \citep{Boothroyd:1988du,2002PASA...19..515K}. Overshoot into the CO core could take place during a late helium shell flash, which is when the post-AGB star becomes helium-rich, or alternatively might occur during every thermal pulse on the AGB.}

We find general agreement between the neon and silicon abundances, which match to within 10\% and 40\%, respectively. However, our model displays a large overabundance of phosphorus by 240\%, and an even larger overabundance of sulphur, for which our model prediction is  44 times too high to match the observation. Our final intershell sulphur abundances are consistent with the models of Herwig that have no extra mixing into the C-O core during thermal pulses, which predict an abundance of 0.9 times solar \citep{2006PASP..118..183W}.

\section{Discussion \& Conclusions}\label{sec:discussion}

The main finding of this study is that variation in the uncertainties
that affect nucleosynthesis of AGB stars  has little impact on the
abundance of sulphur in AGB models. The uncertainties that we have explored include 
those associated with the nuclear network (e.g., neutron capture
cross sections) and those that deal with the formation of a $^{13}$C-rich region in the He-intershell of AGB models.

In terms of the nuclear network, we have measured the effect of rate uncertainties of the two most important neutron-producing reactions by independently testing the high and low rates for $^{13}$C($\alpha$,n)$^{16}$O from \citet[][NACRE collaboration]{1999NuPhA.656....3A} and high, low, and recommended rates for $^{22}$Ne($\alpha$,n)$^{25}$Mg from \citet{2010NuPhA.841...31I}. Table \ref{tbl:neutron-production-surface-results} shows that variations in these rates within the quoted uncertainties have little effect on sulphur.
The rate of the $^{22}$Ne($\alpha$,n)$^{25}$Mg reaction is highly
uncertain, especially at the temperatures found in the He-shells
of AGB stars \citep[e.g., see discussions
in][]{Longland:2009dp,2012PhRvC..85f5809L,Wiescher:2012is}. At the temperatures 
of $T \lesssim 0.30$ GK found in the He-shell of low-mass
AGB stars, this reaction is only marginally activated. However the
uncertainties quoted at this temperature are considerable, for
example, the rate given by \citet{2010NuPhA.841...31I} varies by a factor of 1.24 at 0.300~GK,
although in comparison the NACRE rate varies by a factor of 47
between the upper limit and the recommended values at the same temperature. In contrast, the NACRE rate
for the $^{13}$C($\alpha$,n)$^{16}$O reaction has smaller quoted uncertainties of only about 30\% 
at temperatures up to $0.15$ GK \citep[][NACRE collaboration]{1999NuPhA.656....3A}. Note that this
reaction is ignited in the He-intershell between pulses, when the temperature
is much lower than during convective thermal pulses.
The rate for $^{13}$C($\alpha$,n)$^{16}$O was recently re-determined by \citet{Pellegriti:2008ii} and \citet{Guo:2012ck}.
The uncertainties quoted in \citet{Guo:2012ck} are now even lower, at $\lesssim 20$\% at 
$1.0 \times 10^{8}$ K.

The temperature at the base of the He-flash convection zone is dependent on the numerical details of the AGB model and
in particular on the treatment of the flash-driven convective boundaries \citep[e.g.,][]{Herwig:2000ua}. 
The inclusion of convective overshoot at the inner boundary of the pulse-driven convective zone (PDCZ)
will transport additional $^{4}$He downward to higher temperatures where it burns 
via the triple-$\alpha$ reaction, leading to increased temperatures in the He shell. 
In a 3 M$_\odot$, $Z=0.02$ model, \citet{Herwig:2000ua} found that the inclusion of diffusive 
convective overshooting (with overshoot parameter $f=0.016$) increased the maximum temperature 
at the base of the PDCZ by 13\% from 0.24 GK to 0.27 GK
\citep[see also Fig.~3 in][]{Lugaro:2003ew}. The application of diffusive overshoot to the base of the PDCZ can cause convective mixing to reach into the degenerate C-O core, however this phenomenon is not seen in 3D hydrodynamic simulations of the core-intershell boundary by \citet{2011ApJ...742..121S}. Note that Herwig found modest sulphur depletion
in the He-shell of his model, where the sulphur abundance was between 0.6--0.9 times the
solar abundance. The sulphur depletion is directly related to the amount of 
convective overshoot (and therefore presumably the He-shell temperature). 
For example, Herwig's most recent models calculated with the MESA code \citep{2011ApJS..192....3P} and the NuGrid Multi-zone Post-Processing Network tool \citep{Herwig:2008ur,Bennett:2012gj} and with a lower overshoot parameter applied to the
PDCZ of $f = 0.008$, still show a very small sulphur depletion at about 0.8 times
the solar abundance.

Our only reproduction of significant sulphur depletions in the He-intershell 
of our 3M$_\odot$, $Z=0.01$ model resulted from significantly increasing the rate of the $^{22}$Ne($\alpha$,n)$^{25}$Mg reaction.
By boosting the rate of $^{22}$Ne($\alpha$,n)$^{25}$Mg by a factor of 600, we obtain
a sulphur intershell abundance that is 0.47 times the solar abundance (but only $\approx 0.7$ times
the initial abundance) as shown in Table \ref{tbl:intershell-observations}.  As mentioned in Section \ref{sec:nucmodelresults}, the corresponding surface abundance predictions relative to our standard rate case include: a reduction in 
Ne (0.3 dex), an increase in Mg (0.5 dex), an increase in P (0.4 dex), and only a 
small decrease in S (0.01 dex). 
Note that the factor of 600 is approximately equivalent to a 33\% increase in the maximum He-shell temperature
from 0.30 GK to 0.40 GK, where the upper-limit of the \citet{2010NuPhA.841...31I} rate
increases from $3.01 \times 10^{-11}$ to $1.80 \times 10^{-8}$ cm$^3\cdot$g$^{-1}\cdot$s$^{-1}$, i.e., 
a factor of 631. An increase to the He-burning shell temperature of 33\% is well beyond the 13\% increase predicted by \citet{Herwig:2000ua} with the inclusion of diffusive convective overshoot. Furthermore, an increase to the reaction rate of this magnitude is well outside of experimental uncertainties, indicating the difficulty in depleting sulphur via neutron captures in AGB models.

In our investigation of the uncertainties related to $^{13}$C pocket formation, we found that the insertion of a partially mixed zone ($10^{-3}$ M$_\odot$ and larger) caused a very small reduction in the final intershell sulphur abundance by about 5-10\% and that the use of larger PMZ masses does not necessarily result in greater sulphur depletions (Table \ref{tbl:intershell-observations}). This is because neutron captures at the top of the intershell produce sulphur via phosphorus decay at approximately the same rate as they deplete it via decay into chlorine. However, because the neutron capture rate of $^{30}$Si decreases as a function of temperature while those of phosphorus and sulphur isotopes increase, activation of the $^{22}$Ne($\alpha$,n)$^{25}$Mg reaction during convective thermal pulses at high temperature causes some depletion of phosphorus and sulphur.

Models in the narrow mass range of $2.5$ to $3.5$ M$_\odot$ are known
to efficiently produce neon and increase their Ne/O ratio during AGB
evolution \citep{2003PASA...20..393K}. The elemental increase in neon
is due to $^{22}$Ne production from primary $^{14}$N in
convective pulses via the reaction chain
$^{14}$N($\alpha$,$\gamma$)$^{18}$F($\beta^+$)$^{18}$O($\alpha$,$\gamma$)$^{22}$Ne.
We find that the inclusion of a partial mixing zone significantly
increases the final surface abundance of neon, 
and that larger PMZ sizes generally result in higher neon abundances (Table
\ref{tbl:rates-surface-results}). The extent in mass of the PMZ correlates with
the size of the resulting $^{14}$N pocket, which contributes to the
quantity of $^{14}$N available for the production of $^{22}$Ne.

The positions of our models with large PMZ sizes in the Ar versus Ne plane are far from the observational PNe data (Figure \ref{fig:surfresults-arvsne}) and from this we conclude that the partial mixing zone masses larger than $5 \times 10^{-3}$ M$_\odot$ (50\% of the He-intershell mass in the 3 M$_\odot$ model) are excluded by the observations.
However, this result is dependent on the uncertain physics associated
with the formation of partially mixed zones and $^{13}$C pockets. For
example, when we inserted the protons into the post-processing code, we
made the choice that the proton abundance decreases exponentially, i.e., linearly in a logarithmic scale. 
Studies of the formation of the $^{13}$C pocket have found profiles that can slightly differ from 
this basic assumption, as well as from each other \citep[e.g., see discussion in][]{2012ApJ...747....2L}.

To explain the abundances of young open cluster AGB stars, \citet{Maiorca:2012dk} find that the effective
$^{13}$C required for the $s$-process in low-mass AGB stars is four times larger in models with $M \lesssim 1.5 $M$_\odot$ than that required
in more massive AGB stars \citep[but see the discussion in][]{DOrazi:2012eu}.  \citet{2012ApJ...746...20K} find that PMZ masses of $1.2 \times 10^{-2}$ M$_\odot$ or greater improve the fitting of AGB models to the C/O ratios and [F/Fe] abundances of stars in the Magellanic cluster NGC 1846. Kamath et al.'s best fitting PMZ mass represented a fraction of 80\% of their 1.86 M$_\odot$ model's $1.5 \times 10^{-2}$ M$_\odot$ intershell, with the large PMZ required to produce enough fluorine to match the observations. Our results for neon and argon in planetary nebulae data are not consistent with such a large partial mixing zone. The conflicting results of PMZ studies demonstrate the need for further work to identify the formation mechanism of $^{13}$C pockets in AGB stars.

In general, AGB models are a good match to fluorine abundances in post-AGB stars \citep[e.g.,][]{2005A&A...433..641W}, so the overproduction of fluorine in our 1.8 M$_\odot$ model relative to PG1159-035 may be indication that a lower initial mass is needed to model the star more accurately.

Although it was not the original aim of this investigation, our finding that neon abundances in PNe can be used to constrain the size of the $^{13}$C pocket in low-mass stars is an important and unexpected result. Other investigations of $^{13}$C pocket size in the literature exist \citep[e.g.,][]{BonacicMarinovic:2007hw}, but our method is independent from the uncertainties of AGB s-process abundances. To date, the extent and profile of the $^{13}$C pocket are still highly uncertain, so the constraint set by neon abundances represents an additional clue to understanding this critical step in heavy element production via the s-process. A wider study with a finer grid of PMZ sizes and more accurate PNe data would likely find a smaller upper bound for the PMZ mass and could result in other, new constraints on mixing processes in AGB stars.

\hl{Our results show that variations within the known uncertainties of nuclear reaction rates and partial mixing zone masses are insufficient to reproduce the sulphur anomalies in PNe and PG1159-035 via low-mass stellar nucleosynthesis models. We conclude that our present knowledge of AGB stellar evolution and the relevant reaction rates does not support an explanation for the sulphur anomaly in terms of the nucleosynthesis in PN-progenitor AGB stars.}

\section*{Acknowledgments}
Both of the authors thank the anonymous referee for helpful comments that have improved this paper. We also thank Richard Stancliffe for detailed discussions and comments on the manuscript, Harriet Dinerstein for useful discussions, Marco Pignatari and Falk Herwig for providing unpublished results, and Louise Howes for comments and corrections.
This work was supported by the NCI National Facility at the ANU. A.I.K. thanks the ARC for support through a Future Fellowship (FT110100475). This research has made use of NASA's Astrophysics Data System.

\bibliography{references}
\bibliographystyle{mn2e}

\bsp

\label{lastpage}

\end{document}